\documentstyle[preprint,aps,eqsecnum,epsf,floats]{revtex}
\begin{document}
\tighten

\def\bfl{{\bbox \ell}}

\newcommand{\gsim}{\raisebox{-0.7ex}{$\stackrel{\textstyle >}{\sim}$ }}
\newcommand{\lsim}{\raisebox{-0.7ex}{$\stackrel{\textstyle <}{\sim}$ }}

\def\bull{\vrule height .9ex width .8ex depth -.1ex}
\def\MeV{{\rm MeV}}
\def\GeV{{\rm GeV}}
\def\Tr{{\rm Tr\,}}
\def\nrcpt{NR\raise.4ex\hbox{$\chi$}PT\ }
\def\ket#1{\vert#1\rangle}
\def\bra#1{\langle#1\vert}
\def\ltap{\ \raise.3ex\hbox{$<$\kern-.75em\lower1ex\hbox{$\sim$}}\ }
\def\gtap{\ \raise.3ex\hbox{$>$\kern-.75em\lower1ex\hbox{$\sim$}}\ }
\def\abs#1{\left| #1\right|}
\def\CA{{\cal A}}
\def\CC{{\cal C}}
\def\CD{{\cal D}}
\def\CE{{\cal E}}
\def\CL{{\cal L}}
\def\CO{{\cal O}}
\def\CZ{{\cal Z}}
\def\bvert{\Bigl\vert\Bigr.}
\def\pds{{\it PDS}\ }
\def\ms{MS}
\def\ddq{{{\rm d}^dq \over (2\pi)^d}\,}
\def\ddqm{{{\rm d}^{d-1}{\bf q} \over (2\pi)^{d-1}}\,}
\def\bfq{{\bf q}}
\def\bfk{{\bf k}}
\def\bfp{{\bf p}}
\def\bfpp{{\bf p '}}
\def\bfr{{\bf r}}
\def\dtr{{\rm d}^3\bfr\,}
\def\bfx{{\bf x}}
\def\dtx{{\rm d}^3\bfx\,}
\def\dfx{{\rm d}^4 x\,}
\def\bfy{{\bf y}}
\def\dty{{\rm d}^3\bfy\,}
\def\dfy{{\rm d}^4 y\,}
\def\dfq{{{\rm d}^4 q\over (2\pi)^4}\,}
\def\dfk{{{\rm d}^4 k\over (2\pi)^4}\,}
\def\dfl{{{\rm d}^4 \ell\over (2\pi)^4}\,}
\def\dtq{{{\rm d}^3 {\bf q}\over (2\pi)^3}\,}
\def\dtk{{{\rm d}^3 {\bf k}\over (2\pi)^3}\,}
\def\dtl{{{\rm d}^3 {\bfl}\over (2\pi)^3}\,}
\def\dt{{\rm d}t\,}
\def\frac#1#2{{\textstyle{#1\over#2}}}
\def\darr#1{\raise1.5ex\hbox{$\leftrightarrow$}\mkern-16.5mu #1}
\def\){\right)}
\def\({\left( }
\def\]{\right] }
\def\[{\left[ }
\def\si{{}^1\kern-.14em S_0}
\def\siii{{}^3\kern-.14em S_1}
\def\diii{{}^3\kern-.14em D_1}
\def\dtwiii{{}^3\kern-.14em D_2}
\def\dthiii{{}^3\kern-.14em D_3}
\def\pziii{{}^3\kern-.14em P_0}
\def\poiii{{}^3\kern-.14em P_1}
\def\ptiii{{}^3\kern-.14em P_2}
\def\ipi{{}^1\kern-.14em P_1}
\def\idii{{}^1\kern-.14em D_2}
\def\fm{{\rm\ fm}}
\def\MeV{{\rm\ MeV}}
\def\CA{{\cal A}}
\def\Czzm{ {\cal A}_{-1[00]} }
\def\Cttm{{\cal A}_{-1[22]} }
\def\Ctzm{{\cal A}_{-1[20]} }
\def\Cztm{ {\cal A}_{-1[02]} }
\def\Czzz{{\cal A}_{0[00]} }
\def\Cttz{ {\cal A}_{0[22]} }
\def\Ctzz{{\cal A}_{0[20]} }
\def\Cztz{{\cal A}_{0[02]} }

\def\Ames{ A }  

\newcommand{\eqn}[1]{\label{eq:#1}}
\newcommand{\refeq}[1]{(\ref{eq:#1})}
\newcommand{\eq}{eq.~\refeq}
\newcommand{\eqs}[2]{eqs.~(\ref{eq:#1}-\ref{eq:#2})}
\newcommand{\eqsii}[2]{eqs.~(\ref{eq:#1}, \ref{eq:#2})}
\newcommand{\Eq}{Eq.~\refeq}
\newcommand{\Eqs}{Eqs.~\refeq}

\def\Journal#1#2#3#4{{#1} {\bf #2}, #3 (#4)}

\def\NCA{\em Nuovo Cimento}
\def\NIM{\em Nucl. Instrum. Methods}
\def\NIMA{{\em Nucl. Instrum. Methods} A}
\def\NPB{{\em Nucl. Phys.} B}
\def\NPA{{\em Nucl. Phys.} A}
\def\PLB{{\em Phys. Lett.} B}
\def\PRL{\em Phys. Rev. Lett.}
\def\PRD{{\em Phys. Rev.} D}
\def\PRC{{\em Phys. Rev.} C}
\def\PRA{{\em Phys. Rev.} A}
\def\ZPC{{\em Z. Phys.} C}
\def\SJP{{\em Sov. Phys. JETP}}

\def\FBS{{\em Few Body Systems Suppl.}}
\def\IJMP{{\em Int. J. Mod. Phys.} A}
\def\UJP{{\em Ukr. J. of Phys.}}


\def\spol{\alpha_{E0}}
\def\qpol{\alpha_{E2}}
\def\Mspol{\beta_{M0}}
\def\Mqpol{\beta_{M2}}


\preprint{\vbox{
\hbox{ NT@UW-98-25}
\hbox{ DUKE-TH-98-171}
}}
\bigskip
\bigskip
\bigskip
\bigskip

\title{$\gamma$-Deuteron Compton Scattering in Effective Field Theory}
\author{Jiunn-Wei Chen\footnote{\tt jwchen@phys.washington.edu},
  Harald W. Grie{\ss}hammer\footnote{\tt hgrie@phys.washington.edu}}
\address{Department of Physics, University of Washington, Seattle, 
WA 98195-1560, USA }
\author{ Martin J. Savage\footnote{\tt savage@phys.washington.edu}}
\address{Department of Physics, University of Washington, Seattle, 
WA 98195-1560, USA}
\address{and\ Jefferson Lab., 12000 Jefferson Avenue, Newport News, 
Virginia 23606, USA}
\author{R. P. Springer\footnote{\rm  On leave from the Department of Physics, 
  Duke University, Durham NC 27708.
    \ \tt rps@redhook.phys.washington.edu.}}
\address{Institute for Nuclear Theory, University of Washington, Seattle, 
WA 98195-1560, USA  }
\maketitle

\begin{abstract}
The differential cross section for $\gamma$-deuteron Compton scattering
is computed to next-to-leading order (NLO) in an effective field theory
that describes nucleon-nucleon interactions below the pion production threshold.
Contributions at NLO include the nucleon isoscalar electric polarizability
from its $1/m_\pi$ behavior in the chiral limit.
The parameter free prediction of the $\gamma$-deuteron differential
cross section 
at NLO is in good agreement with data.
\end{abstract} 
\vskip 1in

\vfill\eject


\section{Introduction}

The differential cross section for $\gamma$-deuteron Compton scattering has
been measured at  incident photon energies of
$49\ {\rm MeV}$ and $69\ {\rm MeV}$\cite{Lucas}.
At these energies the forward scattering amplitude is dominated 
by the charged  proton  while
contributions from the strong interactions that bind the nucleons
into the deuteron are suppressed  by factors of the
binding energy divided by the photon energy.
Contributions from the structure of the nucleon 
are suppressed by factors of the photon energy divided by 
the chiral symmetry breaking scale.
For finite angle scattering the deuteron size plays an important
role by setting the scale of the variation of form factors.
It has been suggested that a precise measurement of $\gamma$-deuteron
Compton scattering will determine the isoscalar
electric and magnetic  nucleon polarizabilities.
Given precise measurements of the proton polarizabilities this would 
determine the neutron polarizabilities.
However, since
the physics responsible for binding the deuteron has the same origin as the
physics giving rise to the nucleon polarizabilities, the separation between
binding effects and nucleon structure  is unclear.

The nucleon electromagnetic polarizabilities measure the deformation
of the nucleon in external electric and magnetic fields 
(for an overview see \cite{Polwork}).
Combined analyses of  $\gamma$-proton Compton scattering and
photoabsorption sum rules give a proton electric polarizability of
$\alpha _{E,p}=(12.1\pm 0.8\pm 0.5)\times 10^{-4}\ {\rm fm}^3$
and magnetic polarizability of
$\beta _{M,p}=(2.1\mp 0.8\mp 0.5)\times 10^{-4}\ {\rm fm}^3$\cite{ppola,ppolb}.
However, the neutron polarizabilities are
much more difficult to obtain.
Since there is no free neutron target, 
either  $\gamma$-deuteron Compton scattering \cite{Lucas,npolc}
or neutron scattering off heavy nuclei is used to extract the
neutron polarizability\cite{npola,npolb}.
One extraction of the neutron electric polarizability  
from its characteristic influence on the energy dependence of the
neutron-$^{208}$Pb scattering cross section
gave a precise value of
$\alpha _{E,n}=(12.0\pm 1.5\pm 2.0)\times 10^{-4}\ {\rm fm}^3$ \cite{npola}.
A more recent analysis gives a different central value
with large uncertainty,
$\alpha _{E,n}=(0\pm 5)\times 10^{-4}\ {\rm fm}^3$\cite{npolb}.
Meanwhile, quasi-free $\gamma$-deuteron Compton scattering 
$d (\gamma ,\gamma ^{^{\prime }}n)p$
gives a value of $\alpha _{E,n}$ consistent with zero but  also 
with a large uncertainty
$\alpha_{E,n}=(10.7_{-10.7}^{+3.3})\times 10^{-4}\ {\rm fm}^3$\cite{npolc}.
An analysis of $\gamma$-deuteron Compton scatting with a simple potential model
gives values of
$\alpha _{E,n}=(20\pm 6\pm 3)\times 10^{-4} \ {\rm fm}^3$
and $\beta _{M,p}=(10\pm 6\pm 3)\times 10^{-4}\ {\rm fm}^3$\cite{Lucas}.
Calculations with more realistic potentials and interactions were
performed for this process \cite{LLa,WWA} with the 
nucleon isoscalar polarizabilities input rather than treated as free parameters.
A more sophisticated potential model calculation is currently being
performed by Karakowski and Miller\cite{KarMil}.

Recently, significant theoretical
effort\cite{Weinberg1,KoMany,Parka,KSWa,CoKoM,DBK,cohena,Fria,Sa96,LMa,GPLa,Adhik,RBMa,Bvk,aleph,Parkb,Gegelia,steelea,KSW,KSW2,MJStalks,CHa}
has produced an effective field theory with consistent power counting
that describes the low-energy interactions of two nucleons.
The leading contribution to two nucleons scattering in an S-wave
comes from local four-nucleon operators. Contributions from pion exchanges
and from higher derivative operators are suppressed by additional powers of
the external nucleon momentum and by powers of the light quark 
masses\cite{KSW,KSW2}. To
accommodate the unnaturally large scattering lengths in S-wave nucleon
nucleon scattering, fine-tuning is required. 
This fine-tuning can complicate
power counting in the effective field theory, but dimensional regularization
with power divergence subtraction (PDS), described in \cite{KSW}, provides a
consistent power counting scheme. The technique successfully describes the
$NN$ scattering phase shifts up to center-of-mass momenta of
${\bf p} \sim 300\ {\rm MeV}$ per nucleon\cite{KSW} in all partial waves. 
The electromagnetic moments, form factors~\cite{KSW2} and
polarizability~\cite{dpol} of the deuteron  as well as
parity violation in the two-nucleon sector~\cite{PVeft}
have been explored with this new effective field theory.
In this paper we perform a model independent, analytic calculation of
$\gamma$-deuteron Compton scattering up to next-to-leading order (NLO) in the
nucleon-nucleon effective field theory.
The gauge invariant set of pion graphs that gives the dominant contribution to the
electric polarizability of the nucleon contribute to $\gamma$-deuteron Compton
scattering at NLO.  
We will not treat the polarizabilities as parameters,
but instead compute the  $\gamma$-deuteron Compton scattering 
including these pion
loop graphs and show that the 
differential cross section is in good agreement with the data.
This avoids the problem of defining ``nucleon structure''  
in the deuteron bound-state
(i.e., we only compute S-matrix elements) and means that we
do not have to deal with the issue of field redefinitions
(i.e., ``off-shell effects'').


\section{Effective Field Theory for Nucleon-Nucleon Interactions}

The terms in the effective Lagrange density describing the interactions
between nucleons, pions, and photons can be classified by the number of
nucleon fields that appear. It is convenient to write 
\begin{equation}
{\cal L} = {\cal L}_0 + {\cal L}_1 + {\cal L}_2 + \ldots, 
\end{equation}
where ${\cal L}_n$ contains $n$-body nucleon operators.

${\cal L}_0$ is constructed from the photon field $A^\mu = (A^0, {\bf A})$
and the pion fields which are incorporated into an $SU(2)$ matrix, 
\begin{equation}
\Sigma = \exp\left({\frac{\displaystyle
2i\Pi}{\displaystyle f}}\right)\ \ \ ,\qquad \Pi = \left( 
\begin{array}{cc}
\pi^0/\sqrt{2} & \pi^+ \\ 
\pi^- & -\pi^0/\sqrt{2} 
\end{array}
\right) \ \ \ \ , 
\end{equation}
where $f=132\ \MeV$ is the pion decay constant. $\Sigma$ transforms under
the global $SU(2)_L \times SU(2)_R$ chiral and $U(1)_{em}$ gauge symmetries
as 
\begin{equation}
\Sigma \rightarrow L\Sigma R^\dagger, \qquad \Sigma \rightarrow e^{i\alpha
Q_{em}} \Sigma e^{-i\alpha Q_{em}} \ \ \ , 
\end{equation}
where $L\in SU(2)_L$, $R\in SU(2)_R$ and $Q_{em}$ is the charge matrix, 
\begin{equation}
Q_{em} = \left( 
\begin{array}{cc}
1 & 0 \\ 
0 & 0 
\end{array}
\right) \ \ \ . 
\end{equation}
The part of the Lagrange density without nucleon fields is 
\begin{eqnarray} {\cal L}_0 &&= {1\over
2} ({\bf E}^2 - {\bf B}^2)  \ +\ {f^2\over 8} \Tr D_\mu \Sigma D^\mu
\Sigma^\dagger \ +\ {f^2\over 4} \lambda \Tr m_q (\Sigma + \Sigma^\dagger)
\ +\ \ldots \ \ \ \ .  \end{eqnarray} The ellipsis denote operators with
more covariant derivatives $D_\mu$, insertions of the quark mass matrix $m_q
= {\rm diag} (m_u, m_d)$, or factors of the electric and magnetic fields.
The parameter $\lambda$ has dimensions of mass and $m_\pi^2 = \lambda (m_u +
m_d)$. Acting on $\Sigma$, the covariant derivative is 
\begin{equation}
D_\mu \Sigma = \partial_\mu \Sigma + ie [Q_{em},\Sigma] A_\mu \ \ \ . 
\end{equation}

When describing pion-nucleon interactions, it is convenient to introduce the
field $\xi = \exp\left(i \Pi/f\right) = \sqrt{\Sigma}$. Under $SU(2)_L
\times SU(2)_R$ this transformations as 
\begin{equation}
\xi \rightarrow L\xi U^\dagger = U\xi R^\dagger, 
\end{equation}
where $U$ is a complicated nonlinear function of $L,R$, and the pion fields.
Since $U$ depends on the pion fields it has spacetime dependence. The
nucleon fields are introduced in a doublet of spin $1/2$ fields 
\begin{equation}
N = \left({ {p \atop n} }\right) 
\end{equation}
that transforms under the chiral $SU(2)_L \times SU(2)_R$ symmetry as $N
\rightarrow UN$ and under the $U(1)_{em}$ gauge transformation as $N
\rightarrow e^{i\alpha Q_{em}} N$. Acting on nucleon fields, the covariant
derivative is 
\begin{equation}
D_\mu N = (\partial_\mu + V_\mu + ie Q_{em}A_\mu )N \, \, , 
\end{equation}
where 
\begin{eqnarray}
V_\mu &&= {1\over 2} (\xi D_\mu \xi^\dagger + \xi^\dagger D_\mu \xi)\nonumber \\
&& = {1\over 2} (\xi \partial_\mu \xi^\dagger + \xi^\dagger \partial_\mu\xi + ie
A_\mu (\xi^\dagger Q \xi - \xi Q \xi^\dagger)) \, \, .
\end{eqnarray}
The covariant derivative of $N$ transforms in the same way as $N$ under $%
SU(2)_L \times SU(2)_R$ transformations
(i.e. $D_\mu N \rightarrow U D_\mu N$)
and under $U(1)$ gauge transformations
(i.e. $D_\mu N \rightarrow e^{i\alpha Q_{em}} D_\mu N$).

The one-body terms in the Lagrange density are 
\begin{eqnarray}
{\cal L}_1 & = & N^\dagger \left(i D_0 + {{\bf D}^2\over 2M_N}\right) N 
+ {ig_A\over 2} N^\dagger {\bf \sigma} \cdot 
(\xi {\bf D} \xi^\dagger - \xi^{\dagger} {\bf D} \xi)
N\nonumber \\
& + &  {e\over 2M_N} N^\dagger
\left( \kappa_0 + {\kappa_1\over 2} [\xi^\dagger \tau^3\xi
  + \xi \tau^3 \xi^\dagger]\right) {\bf \sigma} \cdot {\bf B} N
\nonumber\\
& + &
2\pi \alpha_E^{(N0)} N^\dagger N {\bf E}^2\ +\ 2\pi \alpha_E^{(N1)} N^\dagger
\tau^3 N {\bf E}^2
\ +\ 
2\pi \beta_M^{(N0)} N^\dagger N {\bf B}^2\ +\ 
2\pi \beta_M^{(N1)} N^\dagger \tau^3 N {\bf B}^2
+ \ldots,
\label{lagone}
\end{eqnarray}
where $\kappa _0={\frac 12}(\kappa _p+\kappa _n)$ and
$\kappa _1={\frac 12}(\kappa _p-\kappa _n)$
are isoscalar and isovector nucleon magnetic moments
in nuclear magnetons, with 
\begin{eqnarray}
\kappa_p & =&  2.79285\ \mu_N \ ,\qquad\kappa_n = - 1.91304\ \mu_N
\ \ \ .
\end{eqnarray}
In using these values of $\kappa_p$ and $\kappa_n$ in eq.~(\ref{lagone})
we have ignored the contribution from pion loops that  scale as
$ m_\pi$ in the chiral limit.
Counterterms contributing to the isoscalar and isovector
electric polarizabilities of the
nucleon are $\alpha _E^{(N0)}$ and $\alpha _E^{(N1)}$
while the corresponding magnetic
quantities are $\beta _M^{(N0)}$ and $\beta _M^{(N1)}$.

The two-body Lagrange density is 
\begin{eqnarray}
\CL_2 &=& -\left(C_0^{(\siii)}+ D_2^{(\siii)} \lambda\Tr m_q\right))
(N^T P_i N)^\dagger(N^T P_i N)
\nonumber\\
 & + & {C_2^{(\siii)}\over 8}
\left[(N^T P_i N)^\dagger
\left(N^T \left[ P_i \overrightarrow {\bf D}^2 +\overleftarrow {\bf D}^2 P_i
    - 2 \overleftarrow {\bf D} P_i \overrightarrow {\bf D} \right] N\right)
 +  h.c.\right]
\nonumber\\
&& + e L_1 (N^\dagger {\bf \sigma} \cdot {\bf B}N) (N^\dagger N) +
e L_2 (N^\dagger {\bf \sigma} \cdot {\bf B} \tau^a N)
(N^\dagger \tau^a N)
\nonumber\\
& + & 2\pi \alpha_4  (N^T P_i N)^\dagger(N^T P_i N) {\bf E}^2
\ +\ 2\pi \beta_4  (N^T P_i N)^\dagger(N^T P_i N) {\bf B}^2
+ \ldots,
\label{lagtwo}
\end{eqnarray}
where $P_i$ is the spin-isospin projector for the spin-triplet channel
appropriate for the deuteron 
\begin{eqnarray}
P_i \equiv {1\over \sqrt{8}} \sigma_2\sigma_i\tau_2
\ \ \ , 
\qquad \Tr P_i^\dagger P_j ={1\over 2} \delta_{ij}
\ \ \ .
\end{eqnarray}
The $\sigma $ matrices act on the nucleon spin indices, while the $\tau $
matrices act on isospin indices. The local operators responsible for $S-D$
mixing do not contribute at either leading or NLO. The $C_0^{(\siii)}$, $%
C_2^{(\siii)}$ and $D_2^{(\siii)}$ coefficients are determined from $NN$
scattering to be 
\begin{eqnarray}
C_0^{(\siii)}(m_\pi) =-5.51\fm^2\ ,\ 
D_2^{(\siii)}(m_\pi) =1.32\fm^4\ ,\ 
C_2^{(\siii)}(m_\pi) =9.91\fm^4
\ ,
\label{eq:numfitc}
\end{eqnarray}
where we have
regulated the divergences of the theory with dimensional regularization and
have 
chosen to renormalize the theory at a scale $\mu =m_\pi $ in
the power divergence subtraction (PDS) scheme\cite{KSW}. 
We note that observables are independent of the choice of
renormalization scale $\mu$.   
Explicit $\mu$ dependence that can appear in amplitudes is
exactly compensated by the renormalization group evolution of coefficients
in the Lagrange density (for a detailed discussion see\cite{KSW}).
A linear
combination of the coefficients $L_{1,2}$ contribute to the magnetic moment
of the deuteron, but neither they nor the coefficients $\alpha _4$ and
$\beta _4$ contribute to $\gamma$-deuteron Compton scattering at the order to
which we are working. In eq.~(\ref{lagtwo}) we have only shown the leading
terms of the expansion in meson fields, namely the terms we need for our
leading plus NLO calculation.


\section{$\gamma$-Deuteron Compton Scattering}

We are interested in the low energy (below pion production threshold) Compton
scattering process
$\gamma (\omega ,{\bf k}) d
\rightarrow \gamma (\omega^{^{\prime }},{\bf k}^{^{\prime }}) d$
where the incident photon of four momentum $(\omega ,{\bf k})$
in the deuteron rest frame (lab frame) scatters off the  deuteron
to an outgoing photon of four momentum
$(\omega ^{^{\prime }},{\bf k}^{^{\prime }})$.
The scattering amplitude can be written in terms of scalar, vector and tensor
form factors corresponding to the $\Delta J=0,1,2$ interactions of the deuteron
field.
The vector and tensor form factors vanish at leading order in the effective
field theory expansion.
As they do not interfere with the scalar form factor in the expression for the
scattering cross section we will neglect them.
In the zero-recoil limit the scattering amplitude for scalar interactions
between the deuteron and the electromagnetic field can be
parametrized by electric and magnetic form factors $F$ and $G$ as
\begin{equation}
\label{fg}
{\cal M}=i\frac{\displaystyle e^2}{\displaystyle M_N}
\left[\ 
  F(\omega ,{\bf\hat{k}}\cdot {\bf \hat{k}^{^{\prime}}}  )\
  {\bf \varepsilon} \cdot {\bf \varepsilon }^{^{\prime }*}
\ +\
G(\omega , {\bf\hat{k}}\cdot {\bf \hat{k}^{^{\prime}}}  )\
({\bf \hat{k}}\times {\bf \varepsilon} )
  \cdot
({\bf \hat{k}}^{^{\prime }}
  \times {\bf \varepsilon }^{^{\prime }*})
\ \right]
\varepsilon_d\cdot\varepsilon_d^{\prime *}
\ \ \  , 
\end{equation}
where $\bf{\hat{k}}$ and $\bf{\hat{k}^{^{\prime}}}$ are unit vectors in the
direction of ${\bf k}$ and ${\bf k}^{^{\prime}}$ respectively, 
${\bf \varepsilon }$ and ${\bf \varepsilon }^{^{\prime }}$ are the
polarization vectors for initial and final state photons,
and $\varepsilon_d$ and $\varepsilon_d^{\prime *}$ are the polarization vectors
of the initial and final state deuterons.
The zero-recoil limit is one in which the initial and final state deuterons
have the same  four velocity, $v_\mu$, and the initial and final state photons
have the same energy $\omega=\omega^\prime$.
Corrections to the zero-recoil limit are suppressed by powers of the photon
energy divided by the deuteron mass.
The functions $F$ and $G$ depend upon the momentum  transfer 
in the scattering due to the large size of the deuteron.

For purposes of power counting we divide
the kinematic range where photon energies $\omega$ are less than
$100\  {\rm MeV}$ into two regimes.
Regime~I covers the range of $\omega $ with $0\leq \omega \lsim B$, and 
regime~II covers the range $B \lsim \omega \lsim 100\ {\rm MeV}$.
Let $Q$ represent the small expansion parameters in the theory.
In regime~I, the photon energy $\omega$ is taken to scale like $Q^2$,
while $m_\pi$ and $\gamma = \sqrt{M_N B}$ scale like $Q$, where $M_N$ is the
nucleon mass, $m_\pi$ is the pion mass, and
$B$ is the deuteron binding energy.
In regime~II, $\omega$ is taken to scale as $Q$.
If a given Feynman diagram does not contain a radiation pion
(an on-shell pion), then in both regimes
each nucleon propagator without external photon momentum flowing through it
scales as $1/Q^2$ and each pion propagator scales as $1/Q^2$. 
However, if the external photon momentum flows through a nucleon propagator,
yielding
$\sim ({\bf q}^2 + \gamma^2 - M_N\omega)^{-1}$, then in regime I it
scales as $Q^{-2}$, while in regime II it has mixed scaling properties, 
i.e., a
combination of $Q^{-2}$ and $Q^{-1}$.
This mixed scaling makes power counting graphs more complicated in
regime II than in regime~I.
In particular, loops involving such propagators are dominated by momenta of
order ${\bf q}\sim \sqrt{M_N\omega}\sim\sqrt{Q}$, and hence the loop
integration $\int d^4q$ scales like $Q^{5/2}$.
However, by retaining this mixed scaling we are able to move smoothly between
regime~I and regime~II.
If a diagram contains a radiation pion,
then in both regimes each loop integration that picks up the
pion pole scales as $Q^4$, and the nucleon
propagators in this loop scale as $1/Q$.

\begin{figure}[t]
\centerline{{\epsfxsize=4.0in \epsfbox{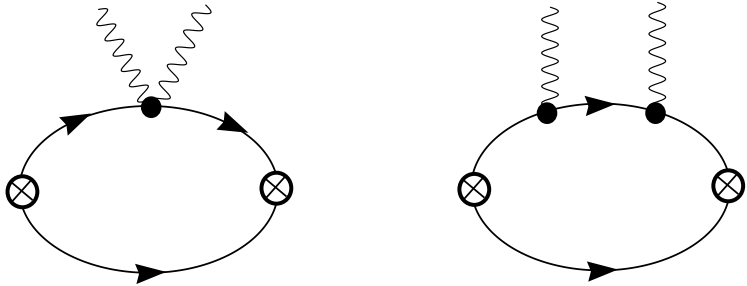}}\hskip 0.8in }   
\vskip 0.1in
\noindent
\caption{\it Leading order contributions to $\gamma$-deuteron
  Compton scattering.
  The crossed circles denote operators that create
  or annihilate
  two nucleons with
  the quantum numbers of the deuteron.
  The dark solid circles correspond to the photon
  coupling via  
  the nucleon kinetic energy operator (minimal coupling).
The solid lines are nucleons.
  The photon crossed graphs are not shown.
}
\label{fig:gdEleclead}
\vskip .2in
\end{figure}

The types of graphs that will contribute to the $\gamma$-deuteron Compton
scattering at LO (starting at $Q^0$) and NLO
(starting at $Q^1$) can be classified as follows:  minimal electric
coupling, $C_2$, potential pion, magnetic
moment coupling, nucleon electric polarizability, and nucleon magnetic
polarizability.
Power counting tells us at what order each of these begins to contribute.  
From eq.~(\ref{lagone}), pion loops contribute to the nucleon 
electric polarizability and scale as $1/m_\pi$~\cite{BKMa,BKKMb,BS,Lvov,HolNa,Hola,BSSa}
at leading order,
contributing to $\gamma$-deuteron Compton scattering at order $Q^1$.
The nucleon magnetic polarizability also receives a contribution that behaves
as $1/m_\pi$ in the chiral limit but is suppressed by a small numerical
coefficient.
It is thought that the nucleon magnetic polarizability will be
dominated by $\Delta$ intermediate states.  
This type of pole graph scales like
$m_\pi^0$ in the chiral limit\cite{BSSa}, and
therefore the magnetic polarizabilities
contribute at order $Q^2$.
The dimension-7 local
polarizability counterterms in eq.~(\ref{lagone}) scale as $Q^2$.
The nucleon electric polarizability contributes to
$\gamma$-deuteron Compton scattering at NLO in regime II,
but is suppressed by two additional powers of $Q$ in regime I.
The $C_2$ operator in eq.~(\ref{lagtwo}) contributes at NLO.

The form factors 
$F$ and $G$ can be expanded in powers of $Q$,
\begin{equation}
\begin{array}{c}
F=F^{LO}+F^{NLO}+...\quad , \\ 
G=G^{LO}+G^{NLO}+...
\ \ \ ,
\end{array}
\end{equation}
It is convenient to introduce related form factors
$\widetilde{F}$ and $\widetilde{G}$ which are related to
$F$ and $G$ by
\begin{eqnarray}
F(\omega ,{\bf\hat{k}}\cdot {\bf \hat{k}^{^{\prime}}})\
 & = & \ \widetilde{F}(\omega ,{\bf\hat{k}}\cdot {\bf \hat{k}^{^{\prime}}} )\ +
 \widetilde{F}(-\omega ,{\bf\hat{k}}\cdot {\bf \hat{k}^{^{\prime}}} )
 \nonumber\\
G(\omega ,{\bf\hat{k}}\cdot {\bf \hat{k}^{^{\prime}}} )\ & = & \ 
\widetilde{G}(\omega ,{\bf\hat{k}}\cdot {\bf \hat{k}^{^{\prime}}})
\ +\ \widetilde{G}(-\omega ,{\bf\hat{k}}\cdot {\bf \hat{k}^{^{\prime}}})
\ \ \  ,
\end{eqnarray}
which also have expansions in powers of $Q$.

\begin{figure}[t]
\centerline{\epsfxsize=4.0in \epsfbox{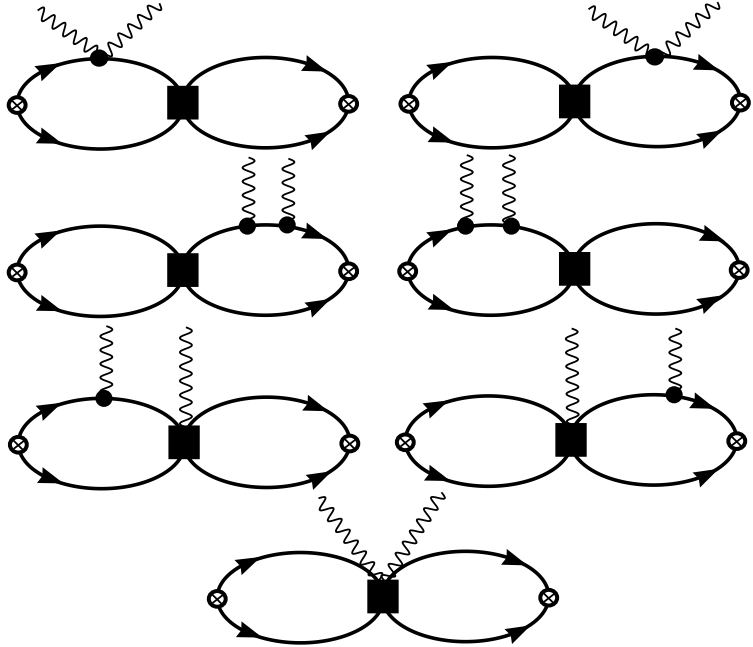}}
\noindent
\caption{\it Graphs from insertions of the operator
with coefficient $C_2(\mu)$ that contribute to
$\gamma$-deuteron Compton scattering at NLO.
  The crossed circles denote operators that create or
  annihilate
  two nucleons with
  the quantum numbers of the deuteron.   
  The solid circles correspond to the photon coupling via
  the nucleon kinetic energy operator (minimal coupling)
  while
  the solid square denotes the $C_2(\mu)$ operator.  
  The solid lines are nucleons.
  Photon cross graphs are not shown.
  }
\label{fig:PolsubC}
\vskip .2in
\end{figure}
%

\begin{figure}[t]
\centerline{\epsfxsize=4.0in \epsfbox{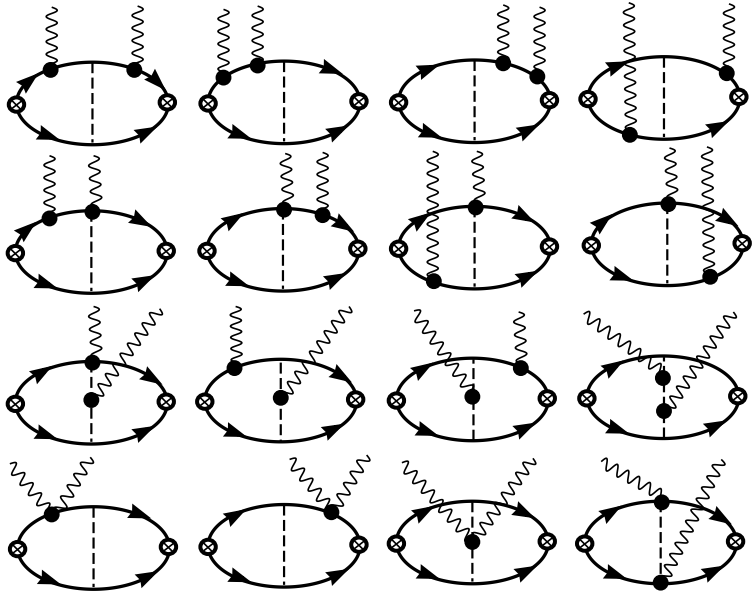}}
\noindent
\caption{\it Graphs from the exchange of a single
  potential pion that
  contribute to $\gamma$-deuteron Compton scattering at NLO.
  The crossed circles denote operators that create or
  annihilate two nucleons with
  the quantum numbers of the deuteron.
  The solid circles correspond to the photon coupling via
  the nucleon or meson kinetic energy operator
  (minimal coupling)
  or from the gauged axial coupling to the meson field.
  Dashed lines
  are mesons and solid lines are nucleons.
  Photon crossed graphs are not shown.
  }
\label{fig:Polsubpi}
\vskip .2in
\end{figure}

\subsection{Regime I}

In regime I (denoted by a subscript on the form factors),
the LO contributions come from the electric coupling
in the
$N^{\dagger }{\bf D}^2N$ operator (minimal coupling),
as shown in Fig.\ \ref{fig:gdEleclead},
\begin{equation}
\widetilde{F}_I^{LO}\ =\widetilde{F}_E\quad , 
\end{equation}
with
\begin{eqnarray}
 \tilde  F_E & = & -\left[
 {\sqrt{2} \gamma\over |\omega | \sqrt{1-{\bf\hat{k}}\cdot {\bf \hat{k}^{^{\prime}}}
     } } \tan^{-1}\left( {|\omega |\sqrt{1-{\bf\hat{k}}\cdot {\bf
         \hat{k}^{^{\prime}}}}\over 2 \sqrt{2} \gamma}\right) 
 \ +\
 {2\gamma^4\over 3 M_N^2\omega^2} \ -\
 {2\over 3} {\gamma (\gamma^2 - M_N\omega - i \epsilon)^{3/2}\over
   M_N^2\omega^2}
 \right]
\ \ \ .
\label{fe}
\end{eqnarray}
In the form factors in eq.~(\ref{fe}) we have neglected terms
suppressed by  additional factors of order
$ {\bf k}^2/M_N\omega$ 
(i.e. recoil effects) compared to the terms presented
(the omitted recoil effects are NNLO in regime II).
This approximation leads to a few percent correction to the rate.
The maximum deviation of the arctangent term in 
eq.~(\ref{fe}) from ${1\over 2}$
is approximately $\sim 15\%$ for $\omega = 100\ {\rm MeV}$, the same
magnitude as formally higher order terms in the $Q$ expansion.

The diagrams shown in Figs.~\ref{fig:PolsubC} and \ref{fig:Polsubpi}
from insertions of the $C_2$ operator and the exchange of a single potential pion
contribute at NLO in regime I.
We write
\begin{equation}
\widetilde{F}_I^{NLO}=\widetilde{F}_{C_2}+\widetilde{F}_\pi \quad , 
\end{equation}
with  the operator with coefficient $C_2$ contributing
\begin{eqnarray}
  \tilde  F_{C2}   = 
 - {C_2(\mu) (\mu-\gamma)^2\gamma\over 3\pi M_N\omega^2}
 & & 
  \left[ \gamma^4 
    - \gamma ( \gamma^2- M_N \omega - i \epsilon)^{3/2}
    \right.\nonumber\\
    & & \left.
      + {3\over 4} M_N^2\omega^2 \left(1-
       { \sqrt{2} \gamma\over |\omega | \sqrt{1-{\bf\hat{k}}\cdot {\bf \hat{k}^{^{\prime}}}
     } } \tan^{-1}\left( {|\omega |\sqrt{1-{\bf\hat{k}}\cdot {\bf
         \hat{k}^{^{\prime}}}}\over 2 \sqrt{2} \gamma}\right) 
\right)
  \right]
\ \ \ .
\label{fc2}
\end{eqnarray}
The exchange of a single potential pion gives a contribution 
independent of the renormalization scale $\mu$, of
\begin{eqnarray}
 \tilde  F_{\pi}  & = &
 {g_A^2 m_\pi\gamma \over 72 \pi f^2 \omega^2 }
 \left[
   {m_\pi (m_\pi^2-2\gamma^2 + 2 M_N \omega)
   \over M_N}
   \log\left[ { m_\pi + 2 \sqrt{\gamma^2 - M_N\omega - i\epsilon}\over \mu}\right] 
\right.\nonumber\\ & & \left.
    \ -\   {2 m_\pi (m_\pi^2-8\gamma^2 + 11 M_N \omega) 
   \over M_N} 
   \log\left[ {m_\pi + \gamma + \sqrt{\gamma^2 - M_N\omega - i\epsilon}\over\mu} \right]
\right.\nonumber\\ & & \left.
  \ +\  {m_\pi (m_\pi^2-14\gamma^2)
   \over M_N}\log\left[ {m_\pi + 2 \gamma\over\mu} \right]
\right.\nonumber\\ & & \left.
\ +\  { 4 \omega (M_N\omega-m_\pi^2) \over m_\pi + \gamma + \sqrt{\gamma^2 -
    M_N\omega - i\epsilon}}
 \ -\  {2 M_N m_\pi\omega^2\over (m_\pi+2\gamma)^2}
\right.\nonumber\\ & & \left.
\ +\  {4 m_\pi \left( 3(\gamma^2- M_N\omega)(\gamma - \sqrt{\gamma^2 - M_N\omega -
      i\epsilon})\right)\over M_N (m_\pi+2\gamma)}
\right.\nonumber\\ & & \left.
\ -\  {2\left(\gamma- \sqrt{\gamma^2 - M_N\omega - i\epsilon}\right)(2 M_N\omega -
  m_\pi\gamma)\over M_N}
\right]
\ \ \ .
\label{fpi}
\end{eqnarray}
In eq.~(\ref{fpi}) we have neglected the finite three-momentum transfer
to the deuteron since it makes a numerically small modification to the amplitude.   
Therefore, we have not presented the complete calculation at 
NLO, but the omitted terms are numerically of order NNLO.

In regime I, the magnetic amplitudes vanish,
\begin{equation}
\widetilde{G}_I^{LO}=0\quad , \\ \widetilde{G}_I^{NLO}\ =0
\quad ,
\end{equation}
 and in the limit of $\omega \ll B,$ the full 
 amplitude can be expanded in powers of $\omega /B$.
The leading term is the Thomson limit for scattering from a charged deuteron
\begin{equation}
{\cal M}=-i\frac{\displaystyle e^2}{\displaystyle 2 M_N}\
{\bf \varepsilon \cdot \varepsilon }^{^{\prime }*}\ 
{\bf \varepsilon_d \cdot \varepsilon_d }^{^{\prime }*}
\approx
-i\frac{\displaystyle e^2}
{\displaystyle M_d}{\bf \varepsilon \cdot \varepsilon }^{^{\prime }*}\ 
{\bf \varepsilon_d \cdot \varepsilon_d }^{^{\prime }*}
\quad , 
\end{equation}
while the  ($\omega /B)^2$ term gives the electric polarizability\cite{dpol}.
Another interesting limit is $B\rightarrow 0$ in which the forward scattering
amplitude reduces to the Thomson limit for scattering from a charged proton,
$-ie^2/M_N$.

\begin{figure}[t]
\centerline{ {\epsfxsize=4.5in \epsfbox{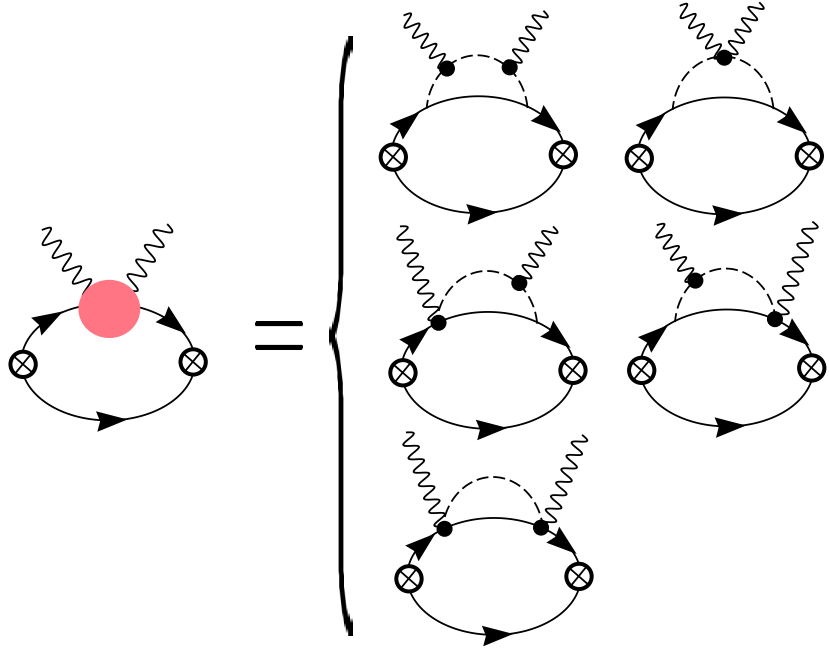}}}
\noindent
\caption{\it Pion graphs that contribute to the nucleon
polarizabilities and to $\gamma$-deuteron Compton scattering at NLO.
  The crossed circles denote operators that create or
  annihilate
  two nucleons with
  the quantum numbers of the deuteron.
  The dark solid circles correspond to the photon
  coupling via the nucleon or pion kinetic energy operator or via the gauged
  axial pion-nucleon interaction.
The solid lines are nucleons and the dashed lines are mesons.
  The photon crossed graphs are not shown.
}
\label{fig:Polariz}
\vskip .2in
\end{figure}

\begin{figure}[t]
\centerline{{\epsfxsize=3.0in \epsfbox{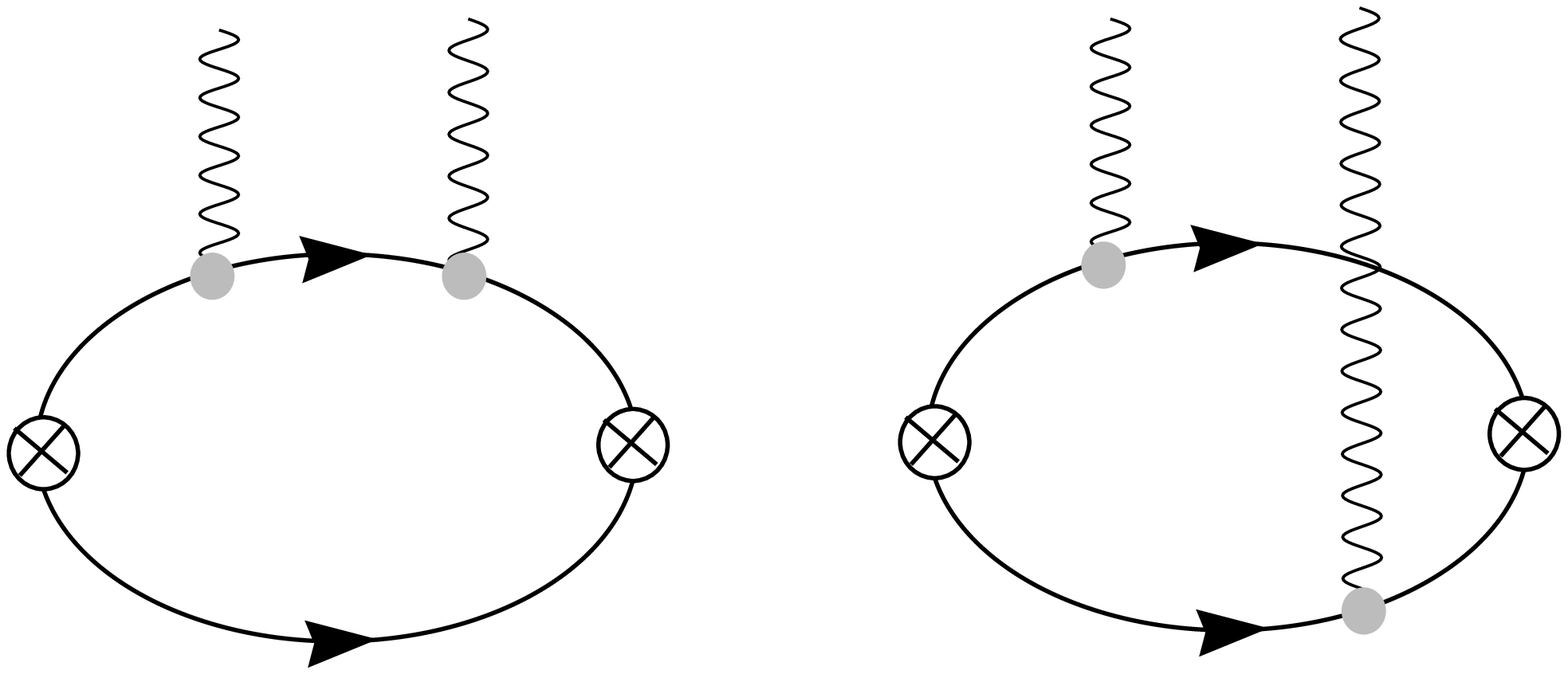}} }
\vskip 0.1in
\centerline{ {\epsfxsize=5in \epsfbox{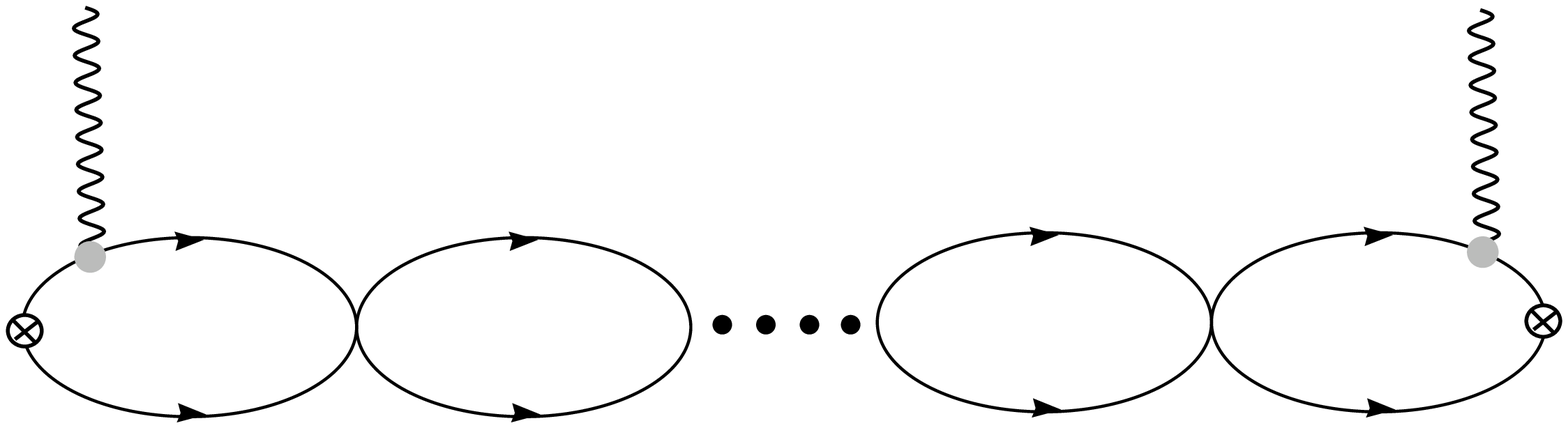}}}
\noindent
\caption{\it Graphs from insertions of the nucleon
  magnetic moment interaction that contribute to $\gamma$-deuteron
  Compton scattering at NLO.
  The crossed circles denote operators that create or
  annihilate
  two nucleons with
  the quantum numbers of the deuteron. 
The light solid circles denote the nucleon magnetic
moment operator.
The solid lines are nucleons.
  The bubble chain arises from insertions of the four
  nucleon operator
  with coefficient $C_0^{(\si)}$ or $C_0^{(\siii)}$ .
}
\label{fig:Maglead}
\vskip .2in
\end{figure}

In regime I, the nucleon polarizability diagrams (Fig.~\ref{fig:Polariz})
and
magnetic moment diagrams (Fig.\ \ref{fig:Maglead}) contribute at NNLO and
N$^3$LO, respectively.
Therefore, in this regime the graphs that give rise to the
nucleon polarizabilities make only a very small contribution
to the polarizabilities of the deuteron.
A more detailed discussion of this point can be found  in ref.~\cite{dpol}.

\subsection{Regime II}

In region II, the seagull diagram shown in Fig.~\ref{fig:gdEleclead}
is LO while the other diagram in Fig.~\ref{fig:gdEleclead}
(and the crossed graph) is demoted to higher order.
At NLO, there are $C_2$ (Fig.\ \ref{fig:PolsubC})
and single potential pion exchange (Fig.\ \ref{fig:Polsubpi}) diagrams, as in
regime I.
The new contributions at NLO are from insertions of the nucleon
magnetic moment interactions
(Fig.\ \ref{fig:Maglead})
and nucleon electric polarizability (Fig.\ \ref{fig:Polariz}) diagrams
(the magnetic polarizability is negligible at this order because of the
 small numerical
coefficient in the pion loop contribution and the counting of the
$\Delta$ intermediate state)
giving
\begin{eqnarray}
\widetilde{F}_{II}^{LO}&&=\widetilde{F}_E\nonumber\, \, , \\
\widetilde{F}_{II}^{NLO}&&=
\widetilde{F}_{C_2}+\widetilde{F}_\pi +\widetilde{F}_{Npol}\nonumber\, \,
, \\ 
\widetilde{G}_{II}^{NO}&&=0\nonumber\, \, , \\
\widetilde{G}_{II}^{NLO}&&=\widetilde{G}_{mag}
\ \ \ \  ,
\end{eqnarray}
where
$\widetilde{F}_E,$ $\widetilde{F}_{C_2}$ and $\widetilde{F}_\pi $ are 
given in eqs. (\ref{fe}) (\ref{fc2}) \& (\ref{fpi}).

The $C_2$ graphs present an interesting issue for power counting in regime II.  
In treating
$M_N \omega$ of order $Q$, we find that individual  $C_2$ graphs contribute at 
$Q^{1/2}$,
$Q^1$, and $Q^{3/2}$.  However, renormalization group invariance for the 
$C_2$ operator
forces the sum of the graphs in the $C_2$ set to contribute at order 
$Q^1$ and higher.

The nucleon electric polarizability contribution at this order
comes from the graphs shown in Fig.~\ref{fig:Polariz}:
\begin{eqnarray}
\widetilde{F}_{Npol}& = &
{ 5 g_A^2 M_N\omega^2\over 48\pi f^2 m_\pi}\ 
\left[ {2 \sqrt{2} \gamma\over |\omega | \sqrt{1-{\bf\hat{k}}\cdot {\bf \hat{k}^{^{\prime}}}
     } } \tan^{-1}\left( {|\omega |\sqrt{1-{\bf\hat{k}}\cdot {\bf
         \hat{k}^{^{\prime}}}}\over 2 \sqrt{2} \gamma}\right)
 \right]
 \left[ 1-{7\over 25} { \omega^2 (1-{\bf\hat{k}}\cdot {\bf
       \hat{k}^{^{\prime}}})\over m_\pi^2} \right]
\ \ \ ,
\label{eq:polfun}
\end{eqnarray}
where an explicit factor of $e^2\over M_N$ has been removed to give the
correct normalization for eq.(\ref{fg}).
The momentum dependence in the ``polarizability'' contribution is generated  
both from
the large size of the deuteron and from the fact that the
pion mass sets the scale of the momentum dependence of electromagnetic
properties of the nucleon.  
The leading momentum dependence
of the pion contribution to
the nucleon polarizability shown in eq.~(\ref{eq:polfun})
has been computed in ref.~\cite{HHKS}.
Formally, all the momentum dependence is the same order in the $Q$ counting, as
$Q\sim m_\pi\sim \omega$.
The set of pion graphs shown in Fig.~\ref{fig:Polariz}
gives an isoscalar nucleon  electric polarizability\cite{BKMa}
of
\begin{eqnarray}
  \alpha_{E,N} & = & {5 g_A^2 e^2\over 192\pi^2 f^2 m_\pi}\ =\ 
1.2\times 10^{-3}\ {\rm fm}^3
\ \ \ \ .
\end{eqnarray}

The magnetic moment contributions, Fig.~(\ref{fig:Maglead}), are
\begin{eqnarray}
\widetilde{G}_{mag} &&= 
\frac{\displaystyle 2(2\kappa ^{(0)^2}+\kappa ^{(1)^2})\gamma (\gamma
-\sqrt{\gamma
^2-M_N\omega -i\epsilon })}{\displaystyle 3M_N^2}\nonumber \\ &&+\ 
\frac{\displaystyle \ \kappa ^{(1)^2}\gamma (\gamma -\sqrt{\gamma
^2-M_N\omega -i\epsilon }
)^2{\cal A}_{-1}^{(^1S_0)}(\omega -B)}{\displaystyle 6\pi M_N}\nonumber \\
&&+\frac{\displaystyle \
\kappa
^{(0)^2}\gamma (\gamma -\sqrt{\gamma ^2-M_N\omega -i\epsilon })^2{\cal A}
_{-1}^{(^3S_1)}(\omega -B)}{\displaystyle 3\pi M_N}\quad , 
\end{eqnarray}
where the leading order nucleon-nucleon scattering amplitude is\cite{KSW,KSW2}
\begin{equation}
{\cal A}_{-1}^{(^1S_0),(^3S_1)}(E)\ =\
\frac{-\displaystyle C_0^{(^1S_0),(^3S_1)}}
{\displaystyle{1}+\displaystyle C_0^{(^1S_0),(^3S_1)}
  \frac{\displaystyle M_N}{\displaystyle 4\pi }
  (\mu -\sqrt{-M_N E-i\varepsilon }\ )}
\ \ \ \ .
\end{equation}
The coefficient $C_0^{(\si)}$ has been determined from nucleon-nucleon
scattering  in the $\si$ channel~\cite{KSW} to be
$C_0^{(\si)}=-3.34\ {\rm fm^2}$ while 
the coefficient $C_0^{(^3S_1)}$ is given in eq.~(\ref{eq:numfitc}).


\begin{figure}[t]
\centerline{\epsfxsize=3.7in \epsfbox{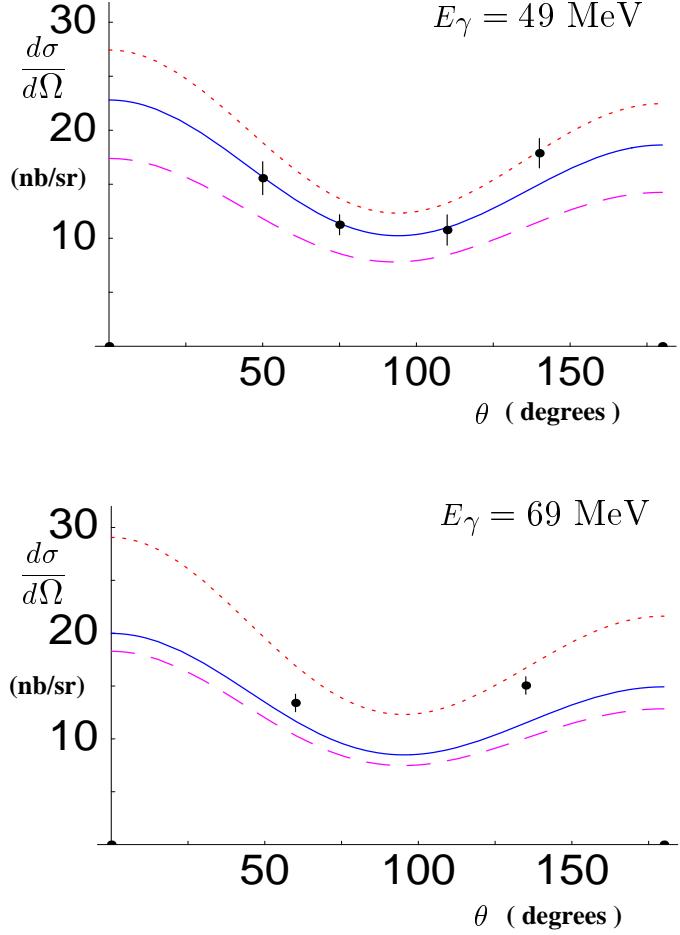}}
\noindent
\caption{\it The differential cross section for
  $\gamma$-deuteron Compton scattering at incident photon
  energies of
  $E_{\gamma}=49\ {\rm MeV}$  and  $69\  {\rm MeV}$.
  The dashed curves correspond to the  LO result.
  The dotted curves correspond to the NLO result without the graphs that
  contribute to the polarizability of the nucleon.
  The solid curves correspond to the complete NLO result
  with no free parameters, as described in the text.
  Systematic and statistical errors associated with each data point 
  have been added in quadrature.
  }
\label{fig:diffcr}
\vskip .2in
\end{figure}

The differential cross section written in terms of the electric and magnetic
form factors in the deuteron rest frame is 
\begin{equation}
\frac{\displaystyle d\sigma }{\displaystyle d\Omega }_{lab}
\ =\
\frac{\displaystyle \alpha^2}{\displaystyle 2 M_N^2}
\left[\ 
\left(\left| F_{II}(\omega )\right|^2
  \ +\ \left| G_{II}(\omega )\right|^2\right)
\left(1\ +\ \cos ^2\theta\right)
 \ +\
 4Re[F_{II}(\omega )G_{II}(\omega )^{*}]\cos \theta
\ \right]
\ \ \ \ ,
\end{equation}
where $\cos\theta ={\bf\hat{k}}\cdot {\bf \hat{k}^{^{\prime}}}$,
the cosine of the angle between the incident and outgoing photons.
Fig.~\ref{fig:diffcr} shows the differential cross section at photon energies
of $49\ {\rm MeV}$ and $69\ {\rm MeV}$.  The dashed curve on both plots is
the LO prediction and is seen to underestimate the observed cross section.
The dotted curve is the differential cross section at NLO but with the omission
of the graphs (shown
in Fig.~\ref{fig:Polariz}) that contribute to the nucleon electric polarizability.  
The solid curve is the parameter free NLO prediction.
We find excellent agreement with the data at $49\ {\rm MeV}$ and reasonable
agreement with the data at $69\ {\rm MeV}$.  
It is clear from Fig.~\ref{fig:diffcr} that omission of the pion graphs
that dominate the nucleon electric polarizability leads to an
over-estimate of the differential cross section, as found in
potential models~\cite{LLa,WWA}.
Further, the agreement between the data and the 
NLO calculation is comparable to the agreement between the data and 
potential model calculations~\cite{LLa,WWA}.
We estimate the size of the NNLO contributions to our calculation of 
the cross section to be about $10\%$, comparable to the experimental 
error associated with each data point.
The analysis of \cite{WWA} agrees well for the $69\ {\rm MeV}$ data and less
well for the $49\ {\rm MeV}$ data while the analysis of \cite{LLa} agrees
reasonably well at both energies.
When the effective field theory calculation is carried out to NNLO
the prediction should be accurate to within $3\%$.

\begin{figure}[t]
\centerline{\epsfxsize=3.7in \epsfbox{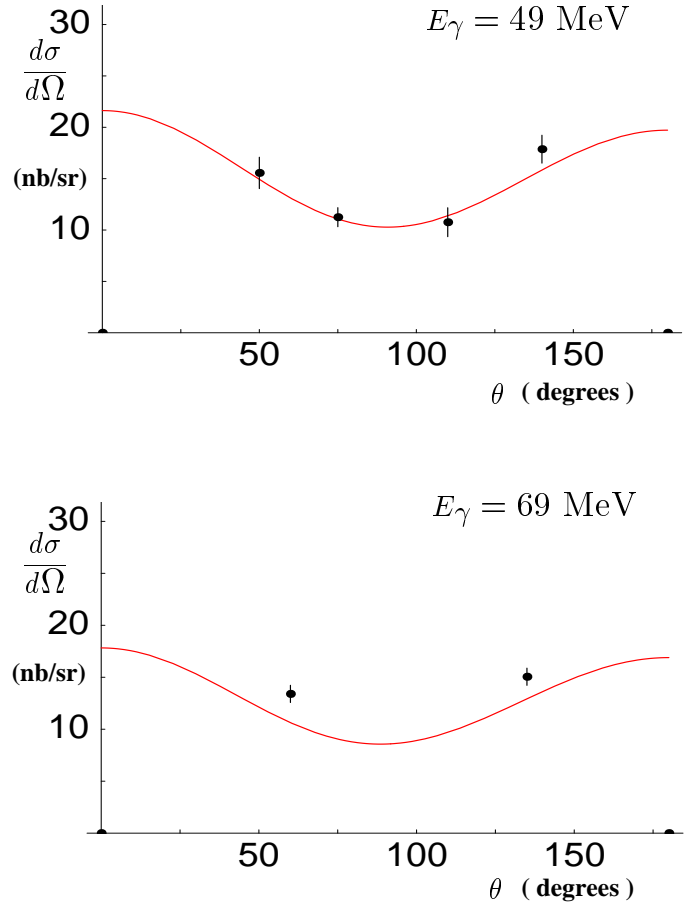}}
\noindent
\caption{\it The differential cross section for
  $\gamma$-deuteron Compton scattering at incident photon
  energies of
  $E_{\gamma}=49\ {\rm MeV}$  and  $69\  {\rm MeV}$.
  The curves correspond to the cross sections at NLO with the  
  nucleon magnetic polarizability fit to the data.
  Systematic and statistical errors associated with each data point 
  have been added in quadrature.
}
\label{fig:difit}
\vskip .2in
\end{figure}

If we assume that the nucleon magnetic polarizability is the 
largest NNLO contribution and fit it to  data we find
a central value of 
\begin{eqnarray}
\beta_{M,N} & = &  {1\over 2}(\beta_{M,p}+\beta_{M,n})
\ =\ 6.5\times 10^{-4}\ {\rm fm}^3
\ \ \ ,
\end{eqnarray}
but with a large uncertainty.
The resulting differential cross section is shown in
Fig.~\ref{fig:difit}.
Agreement with the data is
improved but it must be stressed that there are many contributions 
at NNLO that 
need to be included before reliable conclusions can be drawn.
Naive estimates of $\beta_{M,N}$ from $\Delta$ intermediate state pole-graphs
suggest $\beta_{M,N}$ could be about
$5\times 10^{-4}\ {\rm fm}^3$\cite{BSSa,Lvovb}.


\section{Conclusions}

We have presented analytic expressions for the $\gamma$-deuteron Compton
scattering amplitude at NLO in an effective field theory expansion.
The parameter-free prediction for the differential cross section at NLO
agrees very
well at an incident photon energy of $49\ {\rm MeV}$ and reasonably well
at $69\ {\rm MeV}$, as can be seen in Fig.~\ref{fig:diffcr}.
We see that pion graphs that dominate the electric polarizability 
of the nucleon are
necessary to improve  agreement with the measured
$\gamma$-deuteron cross section.

The theoretical uncertainty in this calculation comes from the omission of
terms at NNLO and higher in the effective field theory expansion, including the
exchange of two potential pions, the exchange of a single radiation pion,
insertions of the $C_4$ operator, relativistic effects, and the vector
and tensor operators.
These higher order terms could modify the differential cross
section at the $10\%$ level.
Calculation of the NNLO terms is required to be sure that the theory is
reproducing the data at the few percent level.

\vskip 0.5in

We would like to thank David Kaplan,
John Karakowski and Jerry Miller for helpful discussions. RPS thanks
the nuclear theory group at the University of Washington for their
hospitality.
This work is supported in part by the U.S. Dept. of Energy under
grants No. DE-FG03-97ER4014 and DE-FG02-96ER40945, and NSF grant
number 9870475.

\end{document}